# Intriguing Magnetocaloric Effect in Multiferroic $Ba_3RRu_2O_9$ (R=Ho, Gd, Tb, Nd) with Strong 4d-4f Correlations

*M. Kumar[1], S. Ghosh[1], G. Roy[1], E. Kushwaha[1], V. Caignaert[2], W. Prellier[2], S. Majumdar[3], V. Hardy[2], T. Basu[1]**

[1] *Rajiv Gandhi Institute of Petroleum Technology, Jais, Amethi, 229304, India*

[2] *Laboratoire CRISMAT, Universite´ de Caen Normandie, ENSICAEN, CNRS UMR 6508, Normandie Univ., 14000 Caen, France*

[3] *School of Physical Science, Indian Association for the Cultivation of Science, 2A & B Raja S. C. Mullick Road, Jadavpur, Kolkata 700 032, India*

*tathamay.basu@rgipt.ac.in

## Abstract

Despite of strong spin-orbit coupling, the magnetocaloric effect (MCE) is not yet explored in larger-d orbital (4d/5d) system. Here we demonstrate the magnetocaloric effect of a 4d-4f correlated multiferroic system, namely $Ba_3RRu_2O_9$ (R= Ho, Gd, Tb, Nd). The compound $Ba_3HoRu_2O_9$ antiferromagnetically orders at 50 K where both the Ho and Ru-moments order, followed by another phase transition ~ 10 K. Whereas, the compound $Ba_3GdRu_2O_9$ and $Ba_3TbRu_2O_9$ orders at 14.5 and 9.5 K respectively, where the ordering of both R and Ru moments are speculated. Our results reveal robust MCE around low-T magnetic phase transition for all the heavy rare-earth members (Ho, Gd, Tb) in this family. The heavy rare-earth members exhibit an intriguing MCE behavior switching from conventional to non-conventional MCE. The change of entropy under a magnetic field for Ho member of this series is ~6 J/Kg-K for 5T field comparable to other 3d-4f magnetocaloric multiferroic systems. Interestingly, the light R-member, $Ba_3NdRu_2O_9$, orders ferromagnetically below 24 K where Nd-moments order, followed by Ru-ordering below 18 K, exhibits a positive MCE below and above magnetic (FM) ordering. The compelling MCE in this series are attributed to temperature dependent complex spin-reorientations for different R-members arising from 4d-4f correlations and rare-earth anisotropy. Our work provides one step closure to design and control efficient MCE in 4d-4f systems.

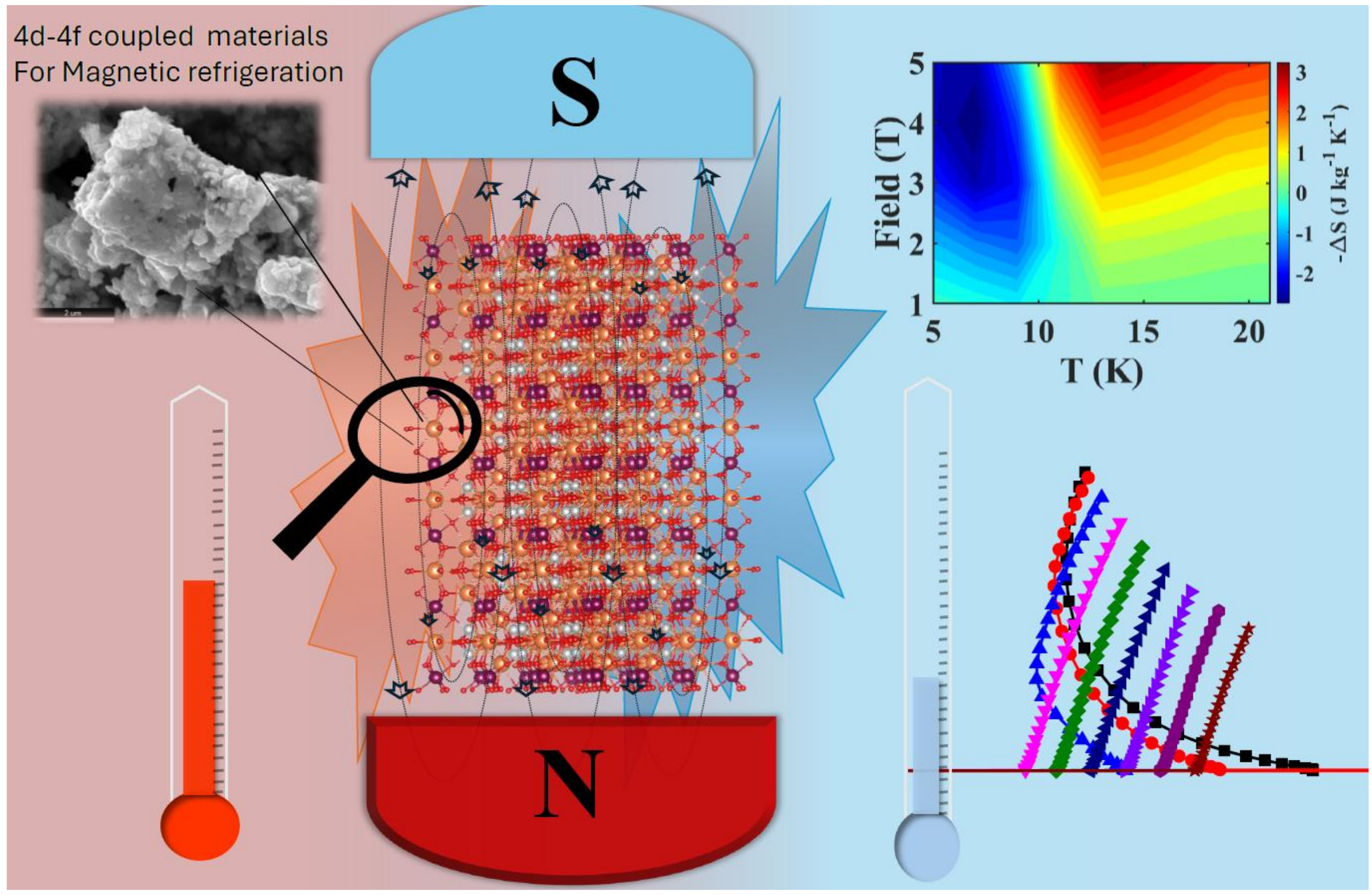

## Introduction

The magnetocaloric effect (MCE) is widely recognized as an environmentally benign and energy-efficient method that has the potential to advance cooling technologies. Apart from its promising future commercial applications of MCE operating at room temperature, magnetic refrigeration emerges as an efficient technology in the space industry and fuel sector as a substitute for cryogenic liquids ($He^3$ systems) in achieving cryogenic temperatures [1-3]. One of clean source of energy is Hydrogen ($H_2$) to use it must have to liquified. The $H_2$ gas is also used as fuel in space shuttles which needs to liquified around 20K, a series of magnetocaloric material is used to achieve from 77K to 20K this cryogenic temperature, as we reporting this rare earth-based MCE materials which can work from whole range of temperature. Eventually, the generation of extremely low temperatures (sub-kelvin) has paramount importance for fundamental research and practical applications, such as, developing new technologies facilitating the operation of quantum computers, developing highly sensitive detectors in astrophysics working at low temperature [4-5]. The magnetic refrigeration of paramagnetic salt is implemented to achieve temperatures below 0.3 K [6-7]. Subsequently, materials with a large magnetic moment and exceptionally low ordering, such as gadolinium gallium garnet $Gd_3Ga_5O_{12}$, were introduced in conjunction with paramagnetic salts [8]. The magnetocaloric effect (MCE) originates from the coupling between the magnetic and thermal degrees of freedom in a material and is fundamentally governed by the change of entropy under an applied magnetic field. In the conventional magnetocaloric effect, adiabatic demagnetization drives a transition from a magnetically ordered to a more disordered spin configuration, resulting in an increase in the change of entropy under a magnetic field that is compensated by a decrease in lattice temperature, thereby producing cooling. Conversely, the inverse magnetocaloric effect, which occurs in certain systems with competing magnetic interactions or complex magnetic phase transitions, is characterized by a counterintuitive response in which adiabatic magnetization leads to a reduction in temperature. In this case, the application of a magnetic field enhances magnetic disorder, increasing the change of entropy under a magnetic field and consequently lowering the sample temperature under adiabatic conditions. It has been observed that the large MCE is influenced by the simultaneous magnetic and structural phase transitions due to spin-lattice coupling, since the applied magnetic field can concurrently alter both magnetic and lattice entropy in these materials [9-11]. The strength of the spin-lattice coupling, often mediated via spin-orbit coupling, is a key parameter in the MCE.

Research on magnetocaloric materials is largely focused on intermetallic systems containing rare-earth (R) ions [12], and magnetically frustrated 3d-4f oxides systems, such as, $RMnO_3$, $RMn_2O_5$, etc [13-18]. In these frustrated oxides, geometrical frustration, strong 3d-4f correlations, and competing exchange interactions between 3d and 4f spins play an important role in the temperature- and magnetic field-dependent spin arrangement/orientation, which releases large change of entropy under variations in the magnetic field, which caused MCE. Although 3d-4f systems have been extensively studied in the search for magnetocaloric effects (MCE), there are rare report of MCE in 4d-4f or 5d-4f systems, despite the fact that 4d/5d systems are currently a focus of research in magnetism due to the competing effects of large spin-orbit coupling (SOC) and crystal electric field (CEF). Hence, 4d-4f (or 5d-4f) systems are considered promising candidates for exhibiting large MCE due to the interplay between large SOC, CEF, and competing exchange interactions, which often results in a magnetically frustrated ground state. As a result, a small change in an external parameter (e.g., magnetic field) may stabilize the ground state and yield large change of entropy under a magnetic field, resulting MCE.

Here we have investigated the magnetocaloric effect on a fascinating strongly correlated Ru(4d)-R(4f) system, 6H- perovskite $Ba_3RRu_2O_9$ (R=rare earth). The compound crystalizes in hexagonal perovskite structure consisting of $Ru_2O_9$ dimer (made of two face-sharing $RuO_6$ octahedra) which are connected by corner-sharing $RO_6$ octahedra [19-25]. The compound $Ba_3NdRu_2O_9$ orders ferromagnetically at around 24 K ($T_C$) where the ordering of only Nd-moments is documented, followed by another magnetic phase transition near 17 K, where the Ru-moments order making a canted AFM structure with Nd. [19,25]. Most likely the less-delocalization of Nd-4f level (which is close to fermi level compared to other heavy R-ions) could be responsible for this FM-behavior of Nd-member in this family. In contrast, all other rare-earth compounds (R= Sm, Gd, Tb, Ho, Er, Dy) have been reported to exhibit antiferromagnetic (AFM) ordering at low temperatures [20-24]. A cooperative magnetic ordering of Ru and Ho through Ru(4d)-Ho(4f) correlation is reported at 50 K ($T_{N1}$) for $Ba_3HoRu_2O_9$, followed by another magnetic phase transition at low temperature ~10 K ($T_{N2}$), characterized as spin-driven ferroelectric phase transition due to strong spin-lattice coupling.

A significant T-dependent spin-reorientation is documented for both for Ru and Ho-members of this family [20-22]. The neutron investigation on $Ba_3HoRu_2O_9$ compound reported intriguing spin-structure with two competing magnetic ground state below $T_{N2}$, implying the presence of strong spin-frustration in this system. In this series, the magnetic rare-earth ions are trivalent ($R^{3+}$) such that the $Ru_2O_9$ dimers have mixed valence $Ru^{4+}/Ru^{5+}$ states [20-21]. The only exception is the $Ba_3TbRu_2O_9$ [21] compound, on which Tb is tetravalent ($Tb^{4+}$), probably due to interplay between spin and lattice degrees of freedom favouring an S-like (orbital moment L=0) magnetic ground state, similar to that of Gd-member ($Gd^{+3}$ possess a S=7/2, L=0 ground state). The compound $Ba_3TbRu_2O_9$ and $Ba_3GdRu_2O_9$ orders antiferromagnetically at 9.5 and 14.5 K respectively [22, 23]. In this dimer series, the Ru-O-Ru bond angle is nearly ~$90^0$ which might favours a weak ferromagnetic exchange coupling through the Ru-O-Ru exchange path, though these dimers are connected via magnetic R-ions or non-magnetic Ba-ions. Whereas the R-O-Ru bond angle is nearly ~$180^0$ which cause a strong antiferromagnetic (AFM) exchange interaction. The dominant magnetic exchange interaction is mediated via Ru-O-R-O-Ru due to strong 4d-4f coupling in this system, which gives rise to complex magnetism and versatile ground states for different R-members. The R-ions are connected through Ru-dimer, therefore, there is negligible direct exchange interactions between magnetic rare-earth ions. However, the lanthanide contraction of R-ions could have pronounced effect on local-distortion of $RuO_6$-octahdera (Ru-Ru distance and change of Ru-O-Ru angle in the dimer), which affects the hybridization of Ru with neighbouring ions and thereby may modify the crystal-electric-field and spin-orbit coupling of Ru-ion. Therefore, compelling effect of magnetic frustration (exchange-frustration), CEF, and SOC, makes this family suitable for investigating the magnetocaloric effect.

**Experimental Techniques:**

The polycrystalline sample of $Ba_3RRu_2O_9$ was prepared by a standard solid-state reaction using mixtures of high purity precursors: $BaCO_3$(Sigma, >99.8%), $RuO_2$(Sigma, 99.9%), and $R_2O_3/R_3O_4$ (Sigma, 99.99%) (R=Ho, Nd, Gd and Tb) which were mixed in an agate mortar and pestle and pressed into pellets, and finally sintered at 1250 °C with several intermediates heating as described in Ref. [24]. The samples have been characterized with X-ray diffraction (Panalytical, EMPYREAN-QTY1) and Scanning Electron Microscopy (JSM-7900F). The details structural analysis is depicted in the **Supplementary Information**.

The heat capacity (C) was measured employing a physical property measurement system (PPMS, Quantum Design) as a function of temperature under various magnetic field. The Magnetization (M) measurement is carried as function of temperature and magnetic field in in a Superconducting Quantum Interreference Devices (SQUID), procured from Quantum design). To estimate the magnetocaloric effect we have calculated the change in entropy ($\Delta S$) of the system using specific heat data across varying magnetic fields using the formula [26,27],

$$\Delta C_p(T)=C_p(T, H) - C_p(T,0)$$

$$\Delta S(0 \rightarrow H, T) = \int_0^{T_o} \frac{\Delta C_p(T')}{T'} dT' + \int_{T_o}^{T} \frac{\Delta C_p(T')}{T'} dT'$$

In the manuscript, the lowest accessible temperature is 2 K for Ho and Nd-members and 5K for R=Tb and Gd-members due to experimental limitations. Therefore, considering the minimal effect of such low temperature, the first theoretical-term in the above equation is ignored. However, we have cross checked the MCE results through magnetization employing Maxwell equation down to 2 K. The change of entropy ($\Delta S$) under applied magnetic field of the system is calculated using below relation:

$$\Delta S = \int_0^H \left\{\frac{\partial M}{\partial T}\right\}_H dH$$

The highest accessible magnetic field is 5T in SQUID magnetometer for magnetic measurements and 9 T in PPMS for heat-capacity measurements. The baseline correction of ($\Delta S$) for 9T magnetic field is done at high temperature normalizing with 5T change of entropy, under magnetic field keeping ($\Delta S$) =0.

One important parameter known as the refrigeration capacity power (RCP) represents the total amount of heat transferred between the hot and cold reservoirs during a single thermodynamic cycle in an ideal refrigeration process. Additionally, the RCP can be calculated using the following equation:

$$RCP = \int_{\mathrm{T_{cold}}}^{\mathrm{T_{hot}}} |\Delta \mathrm{S_{max}}|\, \partial \mathrm{T_{FWHM}}$$

**Result and Discussion**

***$Ba_3HoRu_2O_9$:***

The heat capacity of compound $Ba_3HoRu_2O_9$ under H = 0, 5 T, and 9 T is plotted in Fig. 1a. The C(T) exhibits a sharp peak at 10 K for zero magnetic field, which is suppressed with increasing magnetic field (see left inset of Fig.1a), consistent with the AFM behavior. We did not observe any peak at 50 K; however, a closer look suggests a bifurcation between the curves for different magnetic fields. Fig. 1b depicts the change in entropy under applied magnetic field as a function of temperature for 5 T and 9 T. The long-range-magnetic ordering at 50 K is demonstrated through Neutron diffraction measurements in earlier literatures [20,22]. The absence of peak at 50 K in heat-capacity is due to minimal change of entropy where both the Ho and Ru-moments just start to order with a very small magnetic moment, as described in past literature [20]. The change in entropy under magnetic field increases as the temperature decreases from 50 K, reaching a peak around 10 K, followed by a sharp drop with further decreasing temperature. The maximum change in entropy under magnetic field corresponding to the conventional MCE is 4.6 J/Kg-K around 10 K and then switches to inverse MCE with the maximum value -2.37 J/Kg-K for 9T field while for 5 T field conventional and Inverse MCE is 2.6 J/Kg-K and -2.8 J/Kg-K respectively (Fig. 1b). In order to cross check of large value of change of entropy under magnetic field we have also calculated it from magnetization measurements. Figs. 1c and 1d displays the magnetization of $Ba_3HoRu_2O_9$ from transition temperature regime and change of entropy under magnetic field as function of temperature calculated from magnetization data respectively.

The magnetism below 50 K ($T_{N1}$) is characterized with long-range ordered spin-structure associated with a magnetic wave vector of K= 0.5 0 0. Both Ho and Ru simultaneously order below $T_{N1}$; however, the magnetic moment is very low. The neutron diffraction results show a continuous spin reorientation (along with slow saturation of the moments) as the temperature decreases from 50 K to 12 K, where the canting angle between the Ho and Ru moments slowly changes. Therefore, a sharp change in heat capacity is not observed at 50 K; rather, a slow, gradual change in heat capacity is observed below 50 K down to 12 K, which could result from spin reorientation and the slow increase of magnetic moments as the temperature decreases. A sharp spin reorientation of both Ho and Ru moments within the same spin structure associated with K=0.5 0 0 is documented at $T_{N2}$, along with the emergence of another magnetic phase with a different spin structure, associated with a different wave vector K= 0.25 0.25 0, below $T_{N2}$ [see Ref. 20]. This sharp spin-reorientation and simultaneously emergent of a new magnetic phase results in a large change in entropy under magnetic field, manifesting as a sharp peak in C(T) in the absence of magnetic field. However, the neutron diffraction demonstrated that the emergent of new magnetic phase is not a true-long-range ordered phase, rather development of magnetic domains are speculated [20]. These two low-T magnetic phases compete with each other, striving to stabilize and minimize magnetic frustration. Clearly, external parameters, such as the magnetic field, should have a pronounced effect on this magnetic frustration, which may lead to a large drop in the change in entropy under magnetic field $\Delta S$ below $T_N$. In addition, the occurrence of spin-driven ferroelectricity below $T_{N2}$ indicates strong spin-lattice coupling [22], which contributes to the release of large change in entropy under magnetic field in this compound below $T_{N2}$. The appearance of ferroelectricity will also contribute to the change in entropy which might cause a large-change in heat-capacity at $T_{N2}$. Such a sharp peak in heat-capacity often arises due to $1^{st}$-ordered type phase transition. A H-induced meta-magnetic transition around 4 T ($H_C$) below $T_{N2}$ is reported for this compound in temperature dependent magnetization and dielectric constant behavior [22]. The application of high-magnetic field above $H_C$ can modify the magnetic structure from a non-collinear AFM ordered state to a ferromagnetism-like ordered state. The further enhancement of $\Delta S$ below 8 K (manifesting a dip) in Fig.1b is attributed to the change in entropy under magnetic field resulting from this meta-magnetic phase transition, an ordered-ordered phase transition. The $\Delta S$ tends towards zero at lowest accessible temperature (2 K), which suggest the entropy gets zero at further lower temperature. Right inset of Fig.1a shows an enlarged view of the $C_p$ vs T around 20-25 K. We observe a weak but clear feature ~ 20 K. The neutron diffraction results (see Ref.20) already documented an anomaly around this temperature for the magnetic reflection (0.5 0 2) related to magnetic wave vector k=0.5 0 0, suggesting the sharp reorientation of the moments at this temperature. This reorientation causes the change in entropy which is manifested in an anomaly in our C(T) data.

***$Ba_3GdRu_2O_9$:***

Now, we will discuss the MCE effect of $Ba_3GdRu_2O_9$, which consists of a $Gd^{3+}$ ion, which is considered an S-state ion with an orbital moment L = 0. Therefore, in contrast to the Ho-

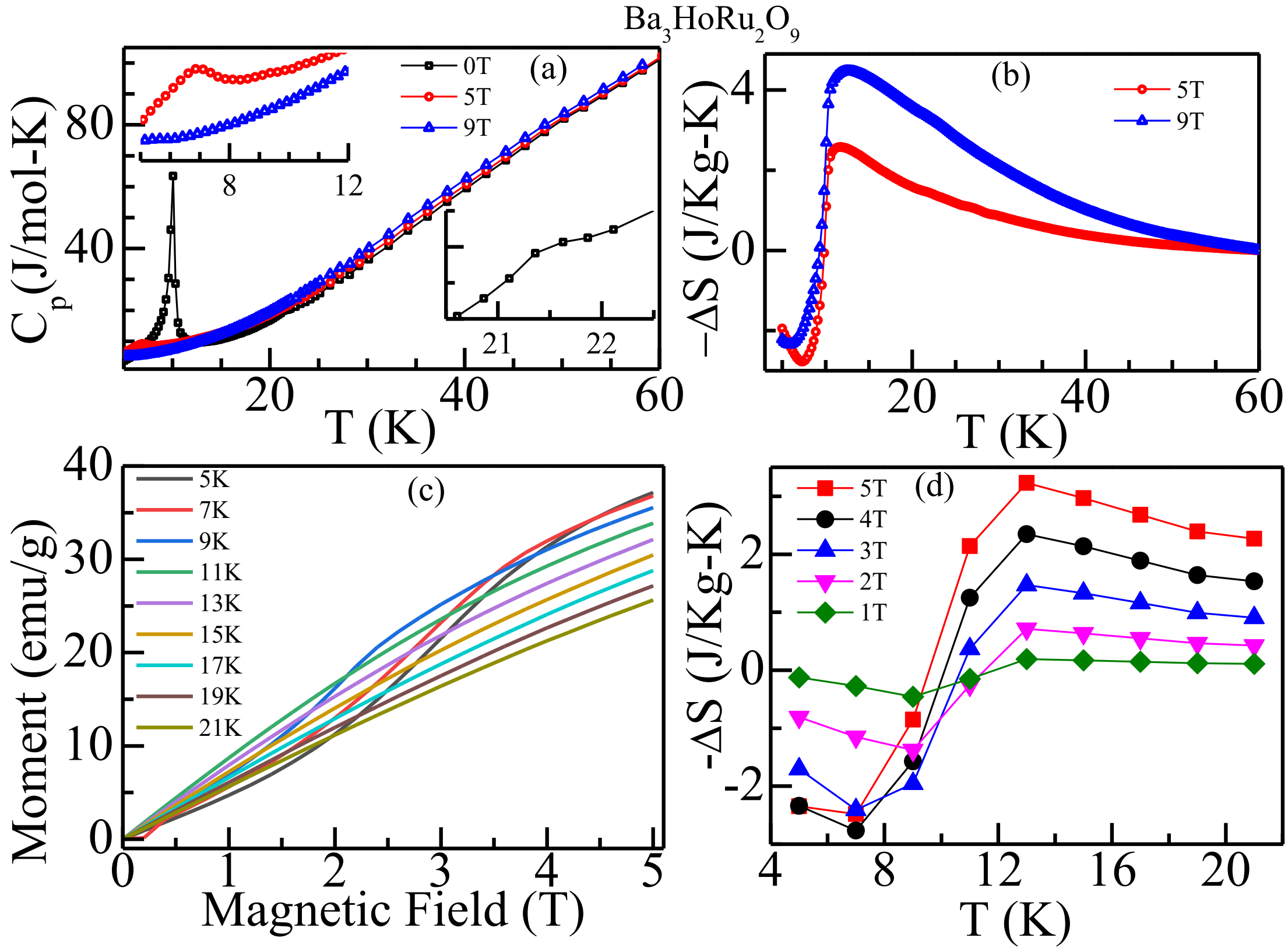

*Figure 1(a)Heat capacity as function of T. Inset of (a) shows the enlarged view of heat-capacity around $T_{N2}$ (b) Change of entropy under magnetic field as function of T. (c)M-H plot (d) Change of entropy under magnetic field as function of T from M-H for $Ba_3HoRu_2O_9$.*

member in this family, Gd does not have significant spin-orbit coupling containing a large spin-moment (S=7/2). However, the spin-orbit coupling of Ru-ion (4d-orbital) should be present. The heat capacity in the absence of the magnetic field reveals a peak at approximately 14.5 K ($T_N$), as depicted in Fig. 2a. Under the application of a magnetic field of 5 T, this peak shifts to lower temperatures, and an additional anomaly emerges near 10 K [23]. With further increases in the magnetic field strength (say H= 9 T), the temperatures corresponding to these two anomalies converge, eventually merging into a single, broad feature that continues to shift to lower temperatures (Fig. 2a). Concurrently, the peak associated with the anomaly diminishes in intensity. These heat capacity results provide clear evidence that the observed transitions represent two distinct phases that coalesce into a single broad anomaly under the influence of the external magnetic field. Fig. 2b illustrates the temperature-dependent ΔS under the application of magnetic fields of 5 T and 9 T. The compound shows a switching from conventional to inverse MCE around 12 K. The maximum entropy change for the inverse MCE is approximately -0.8 J/Kg-K, whereas the conventional MCE reaches a peak value of 1.8 J/Kg-K for 5 T field (Fig. 2b). A conventional to inverse MCE is observed in $GdMn_{1-x}Cr_xO_3$[28]. For application of a higher magnetic field of 9 T, the compound exhibits exclusively conventional MCE, with a maximum entropy change of 4.2 J/Kg-K (Fig. 2b). $Gd^{3+}$ possesses a half-filled 4f shell, resulting in no orbital angular momentum contribution (L = 0) and a pure spin quantum number (S = 7/2). This gives $Gd^{3+}$ a total angular momentum of J = 7/2, leading to very low magnetic anisotropy. Consequently, the energy required to reorient the magnetic moments is minimal, enabling efficient spin alignment with an external magnetic field. The compound shows multiples meta-magentic type phase transition in isothermal M(H) at low temperature [23]. A clear hysteresis around meta-magentic phase transition is observed at 10K which is absent at high-temperature below magnetic ordering [see Ref. 23]. The increase of $\Delta S$ below ~10 K with lowering the temperature is related to this magnetic feature and could be attributed to the enhancement of change in entropy under magnetic field due to a stabilizing different magnetic spin-structure under application of high magnetic field. The value of $\Delta S$ is nearly zero at 5 K. Such behavior could be shifted to a further lower temperature with increasing magnetic field. To confirm that the observed MCE is an intrinsic property of the material and not an artifact, we performed additional verification using magnetization measurements down to 2K. Figs. 2c and 2d show the magnetization of $Ba_3GdRu_2O_9$ near the magnetic transition temperature regime and the temperature-dependent change in entropy

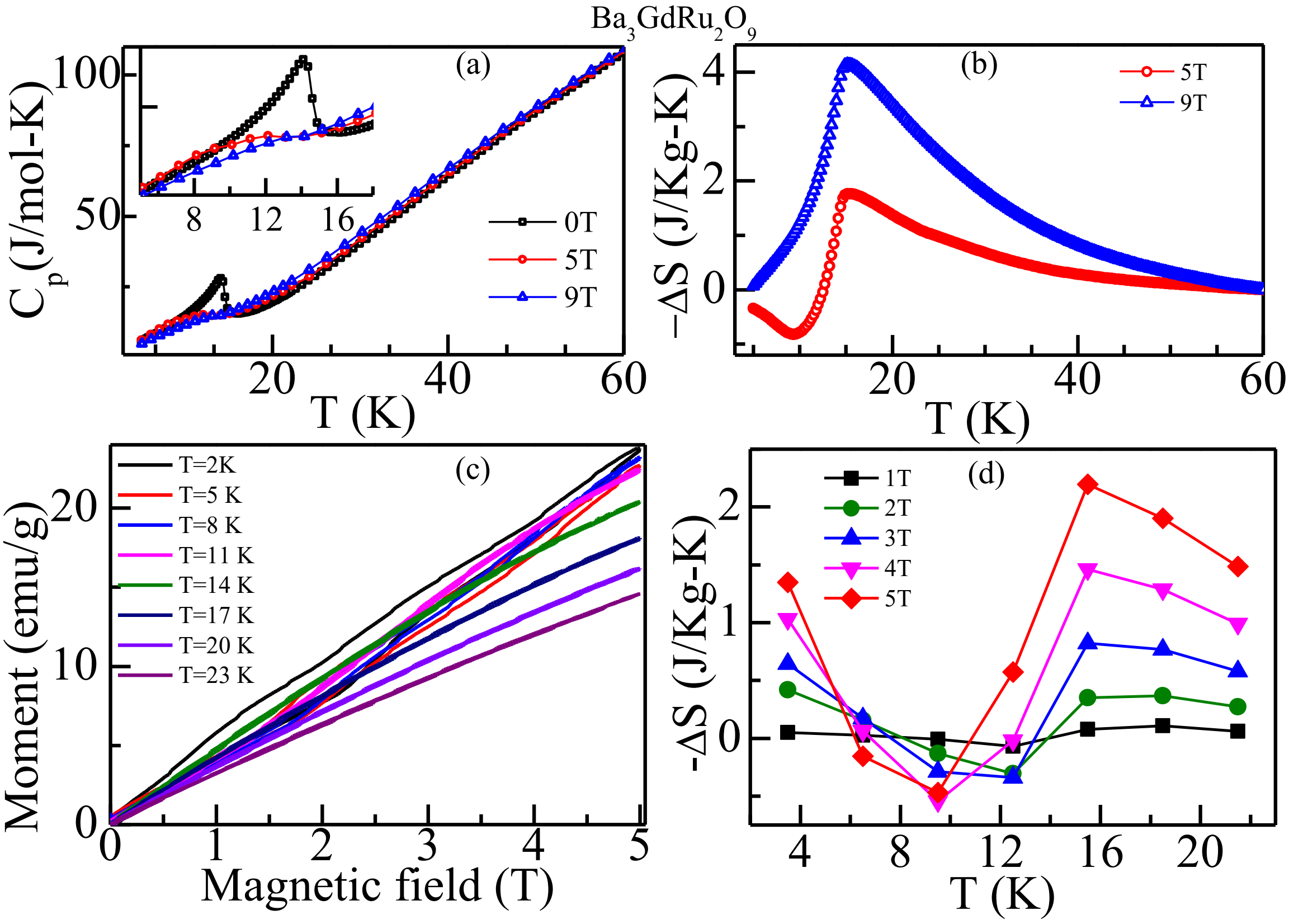

*Figure 2(a)Heat capacity as function of T. Inset of (a) shows the enlarged view of heat-capacity around $T_{N2}$ (b) Change of entropy under magnetic field as function of T. (c)M-H plot (d) Change of entropy under magnetic field as function of T from M-H for $Ba_3GdRu_2O_9$.*

under magnetic field calculated from M-H data as a function of temperature for various applied magnetic fields, respectively. The values of ΔS(T) obtained from magnetization measurements are in excellent agreement with those determined from heat capacity (Figures 2a and 2b). This consistency confirms that $Ba_3GdRu_2O_9$ exhibits inverse MCE at lower temperatures and conventional MCE at higher temperatures, with nearly identical magnitudes for the same applied fields in both measurement techniques.

***$Ba_3TbRu_2O_9$:***

The compound $Ba_3TbRu_2O_9$ undergoes long-range antiferromagnetic ordering at 9.5 K, as evidenced by a sharp peak in the heat capacity data shown in figure 3(a). The application of an external magnetic field suppresses this peak and shifts the ordering temperature to lower values, indicating AFM nature similar to Ho and Tb-members in this series [22]. However, ΔS(T) plot in Fig. 3b elucidate that the system conventional magnetocaloric effect (MCE) is approximately 0.5 J/Kg-K and value of inverse MCE is -1.4 J/Kg-K around $T_N$ under an applied magnetic field of 5 T. Upon increasing the magnetic field to 9 T, the change in entropy under magnetic field enhances further, reaching around 1.5 J/Kg-K and value of inverse MCE

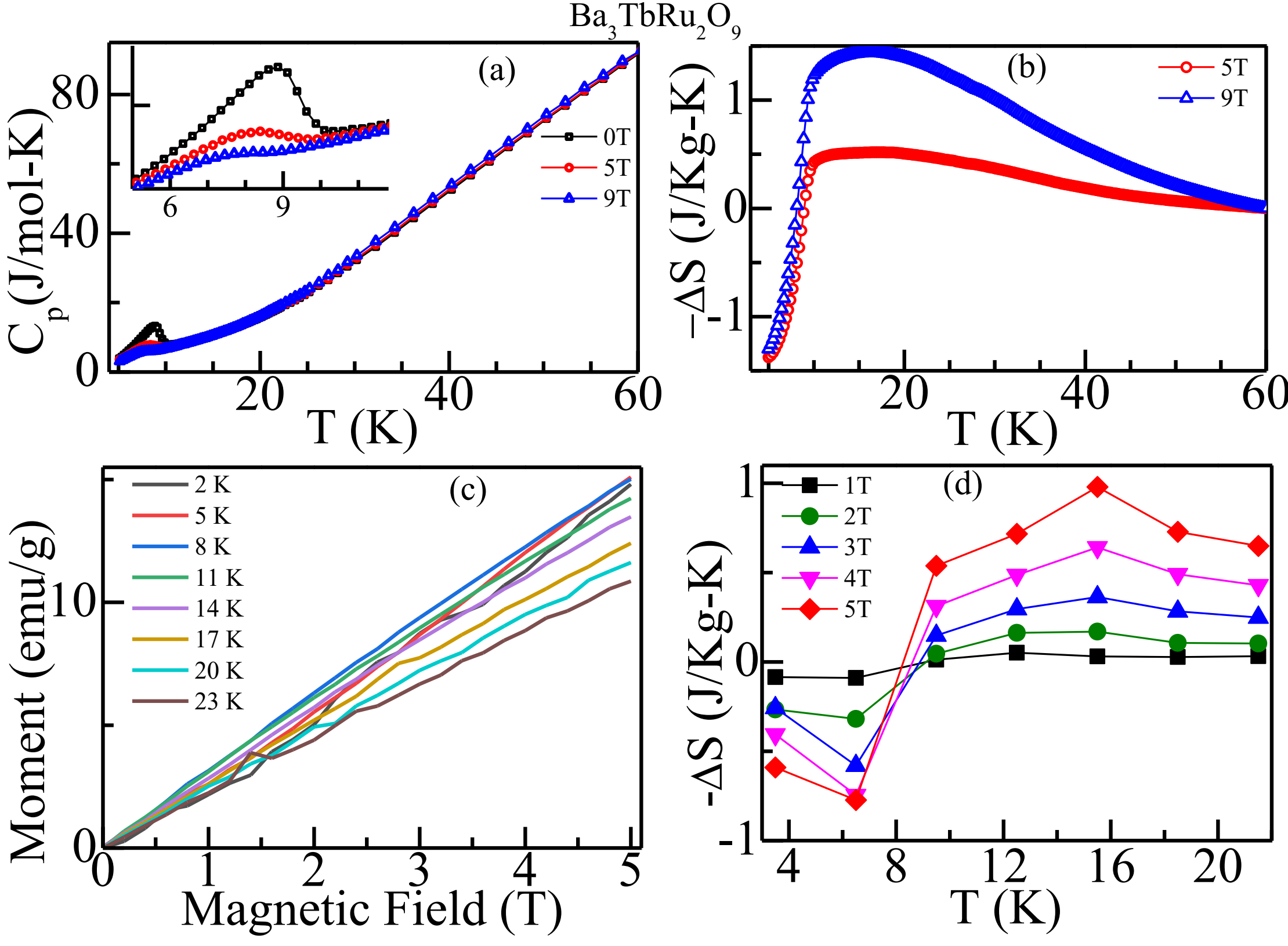

*Figure 3(a)Heat capacity as function of T. Inset of (a) shows the enlarged view of heat-capacity around $T_{N2}$ (b) Change of entropy under magnetic field as function of T. (c)M-H plot (d) Change of entropy under magnetic field as function of T from M-H for $Ba_3TbRu_2O_9$.*

decreasing to -1.3 J/Kg-K. We also observed the same feature through magnetization and verifying it by change in entropy under magnetic field calculation. In this 6H perovskite *d-f* coupled lanthanide series only $Ba_3TbRu_2O_9$ system is showing unique magnetic ground state here. The Tb is showing S-like state showing orbital quenched and Ru showing reduced moment that led to small inverse MCE which is decreasing with increasing field [29]. The heat-capacity results are limited down to 5 K due to experimental resources; however, we have calculated the change in entropy under magnetic field down to 2 K employing Maxwell relations from isothermal M(H) at several temperatures. A clear upturn in change of entropy is observed below ~ 6K, similar to Gd-member in this series. This could be associated with a different spin-structure where $Ba_3TbRu_2O_9$ exhibits a meta-magnetic transition ~ 40 kOe at low temperatures down to 2 K, which is nearly absent at higher temperature ~ 7 K [21]. The nature of MCE is nearly similar for Tb and Gd-members in this family due to similar *s*-orbital like ground state. However, there are distinct difference in the magnetism of these two compounds, namely, the magnetic ordering temperatures and meta-magnetic type transitions at different magnetic field and temperature regime, which is consistent with our MCE results.

***$Ba_3NdRu_2O_9$:***

To understand the role of R-ions, we have explored the MCE of light rare-earth member, $Ba_3NdRu_2O_9$. Fig. 3a illustrates the variation of heat capacity with temperature, revealing a sharp peak at 24 K, for the compound $Ba_3NdRu_2O_9$. Around 24 K, the system undergoes ferromagnetic (FM)-type long-range ordering, where Nd moments become ordered [19]. Fig. 3b shows the change in entropy under magnetic field as a function of temperature. Interestingly, the system exhibits a purely conventional magnetocaloric effect (MCE). Most likely, the FM-nature of this compound is responsible for a pure conventional MCE. We did not observe any significant change in MCE at 17 K, where Ru-moments get ordered [25]. We do not observe inverse MCE or any cross-over for light rare-earth Nd-member in contrast to other heavy rare-earth members. Cooperative magnetic ordering of Ru and R-ions (Ho/Tb), and T-dependent spin reorientation due to strong 4d-4f coupling is reported for these heavy R-members [20, 29]. No such cooperative magnetic ordering and spin reorientation is documented for Nd-members. Hence, the different MCE behavior of $Ba_3NdRu_2O_9$ compared to that of other R-members in this family is consistent with the magnetism. These results suggest that the MCE in $Ba_3NdRu_2O_9$ is primarily dominated by the changes in the change of entropy under magnetic field of R ions. The maximum entropy change for the conventional MCE is 2.4 J/Kg-K at approximately 24 K under an applied magnetic field of 9 T. Interestingly, significant MCE persists even above $T_N$. The earlier study reveals the existence of short-range ordering in this compound even up to 40 K [19]. Eventually, a peculiar shift of the peak in ac susceptibility to higher temperatures is observed under the application of an external dc magnetic field, which indicates a pronounced effect of the magnetic field on spin correlations in this compound above long-range magnetic ordering [ see Ref. 19]. The presence of significant MCE is attributed to a change in entropy resulting from short-range magnetic correlation for this system. The change in entropy is zero nearly at T= 2 K and 40 K. Such a MCE phenomenon above long-range magnetic ordering is rarely observed.

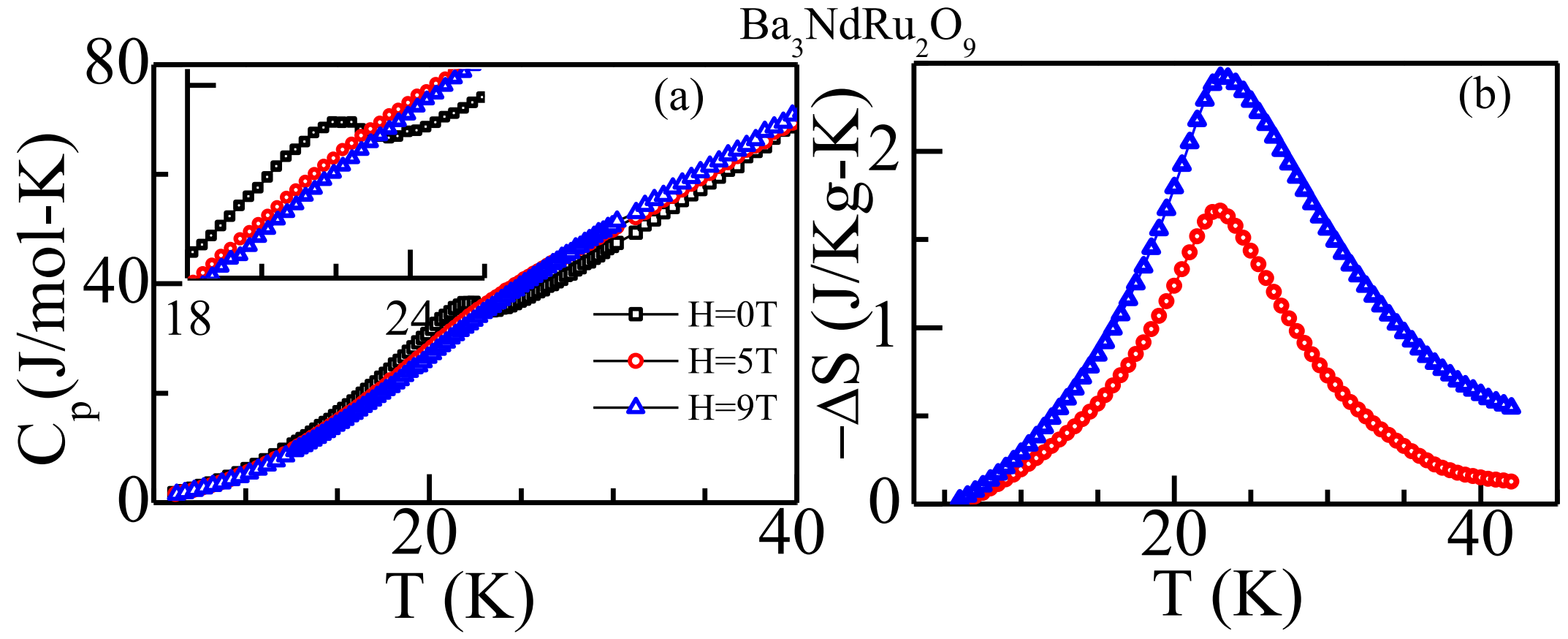

*Figure 4(a) Heat capacity (b) Change in entropy under magnetic field as function of T for $Ba_3NdRu_2O_9$. Inset of (a) shows the enlarged view of heat-capacity around $T_C$.*

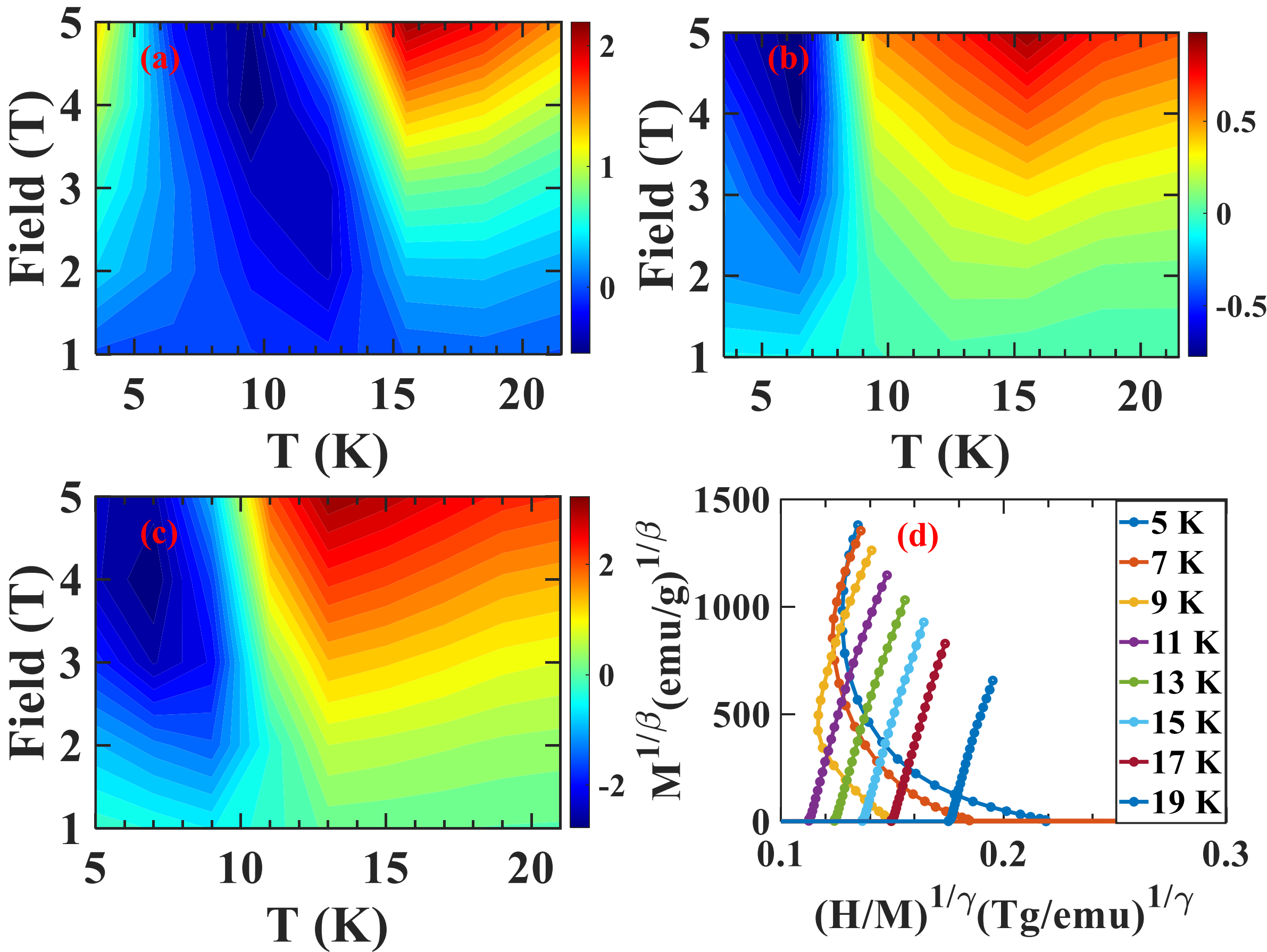

*Figure 5 (a-c) shows the contour plots of the change of entropy under magnetic field ΔS per J/Kg/K as a function of temperature and external magnetic field for Gd, Tb and Ho member for $Ba_3RRu_2O_9$ respectively and (d) shows the Arrott's plot ($M^{1/\beta}$ vs $H/M^{1/\gamma}$) for the Ho member.*

| Compounds | Magnetic ordering T(K) | $-\Delta S_{max}$ J/Kg-K (Field) | | RCP (9T) | references |
|---|---|---|---|---|---|
| | | MCE | Inverse MCE | | |
| $Ba_3HoRu_2O_9$ | 50 K ,10 K | 4.6 (9T) | -2.8(5T) | 86 J/Kg | This work |
| $Ba_3GdRu_2O_9$ | 14.5 K | 4.2 (9T) | -0.8(5T) | 65 J/Kg | This work |
| $Ba_3TbRu_2O_9$ | 9.5 K | 1.5 (9T) | -1.3 (5T) | 34J/Kg | This work |
| $Ba_3NdRu_2O_9$ | 24 K, 17K | 2.8 (9T) | | 38 J/Kg | This work |
| $Y_2Ir_{2-x}Cr_xO_7$(x=0) | 15K | 0.12(7T) | | 49 J/Kg | [11] |
| $Y_2Ir_{2-x}Cr_xO_7$(x=0.05) | 34K | 0.25(7T) | -0.3(7T) | 75 J/Kg | [11] |
| $Y_2Ir_{2-x}Cr_xO_7$(x=0.2) | 35K | 0.68(7T) | -1.73(7T) | 78 J/Kg | [11] |
| $La_{1-x}Na_xMnO_3$(x=0.05) | ~251.5K | 5.77(3T) | | 150 J/Kg | [17] |
| $La_{1-x}Na_xMnO_3$(x=0.1) | ~278.2K | 4.14(3T) | | 160.9 J/Kg | [17] |

*Table 1 Comparison of magnetic ordering, MCE (change in entropy) and RCP*

**Outlook/ Conclusion:**

We have investigated the magnetocaloric effects of a 6H-perovskite multiferroic systems, $Ba_3RRu_2O_9$ (where R=Ho, Gd, Tb and Nd), on which the strong 4d-4f correlation and exchange frustration plays the decisive role to govern MCE. The obtained MCE is comparable to other frustrated multiferroic oxide systems, such as $RMnO_3$, $RMn_2O_5$, etc. [15, 16] Therefore, this 4d-4f correlated systems can be characterized as multiferroic magnetocaloric compound. A contour plot of $dS_M$ as a function of magnetic field and temperature is shown in Fig. 5a-5c for Gd, Tb and Ho member respectively. Interestingly, we observe intriguing MCE in this 4d-4f correlated systems for heavy-rare-earth members (Ho, Gd), that is, a switching from conventional to inverse MCE, unlike other 3d-4f frustrated multiferroic magnetocaloric oxides. Similar switching is reported in some intermetallic magnetocaloric systems [30,31,32]. The switching of MCE is ascribed to the relative spin-orientations of the R-ion with temperature which varies with for different R-ions and strength of 4d-4f coupling, which govern the change in entropy. We have further plotted Arrott's plot for $Ba_3HoRu_2O_9$ compound (see Fig. 5d), where $M^{1/\beta}$ is plotted against $H/M^{1/\gamma}$ and analysed the critical behavior of Ho by optimising a critical exponent for each compound. For Ho ($\beta$=0.5, $\gamma$= 1), from high-field isotherms which shows nearly linear and parallel behavior, consistent with mean-field interactions and that means dominating long-range exchange coupling through weak critical fluctuations. The absence of MCE switching is observed for the compound $Ba_3NdRu_2O_9$, on which the spin reorientations of Nd-moment is not reported. Moreover, we report significant conventional MCE above long-range magnetic ordering for the light-rare-earth member $Ba_3NdRu_2O_9$, attributed to change in change in entropy under magnetic field due to short-range spin-correlation. The magnetodielectric coupling is also reported stronger in heavy-rare earth members (Ho, Tb) which is weak in light rare-earth Nd members [21]. This implies that different degrees of 4d-4f correlations for heavy-rare-earth and light rare-earth members have significant effect on magnetism, magnetodielectric behavior and magnetocaloric behavior. Further microscopic investigations under magnetic fields are warranted to shed light on detail mechanism of this intriguing MCE behavior in this fascinating 4d-4f system, $Ba_3RRu_2O_9$.

**Acknowledgements:**

TB greatly acknowledges the Science and Engineering Research Board (SERB), now called as Anusandhan National Research Foundation (ANRF) (Project No.: SRG/2022/000044), and UGC-DAE Consortium for Scientific Research (CSR) (Project No CRS/2021-22/03/544), Government of India, for funding. WP acknowledge the LAFICS program, supported by the CNRS, and Indo-French collaboration, CEFIPRA. GR acknowledge the Raman-Charpak Fellowship by CEFIPRA. TB Thanks the SSHN fellowship by CEFIPRA, Embassy of France for Indo-French collaboration. MK acknowledges UGC-CSIR for Ph.D. NET fellowship. TB thank the Central Instrumentation Facilities (CIF), RGIPT for sample characterizations through XRD, SEM and TEM. Authors thanks Prof. K. G. Suresh, IIT Bombay, India, and Dr. R. Rawat, UGC-DAE Consortium for Scientific Research, Indore, India, for granting the access of some preliminary experimental facilities.

**Competing interests:**

The authors declare no competing financial interests.